\documentclass[11pt]{atlasnote}
\usepackage{atlasphysics}
\usepackage{subfigure}
\usepackage{mathrsfs}
\usepackage{array}
\usepackage{mcite}
\usepackage{cite}
\usepackage{xspace}
\usepackage{lineno}
\usepackage{url}

\usepackage{xspace}

\newcommand{\ETmiss}{\ensuremath{E_T^{\mathrm{miss}}}}
\newcommand{\aveMu}{\ensuremath{\langle \mu \rangle}}
\newcommand{\us}{\mbox{$\mu$s}}
\newcommand{\invfb}{\mbox{fb$^{-1}$}}

\newcommand{\AxB}[2]{\mbox{{#1}$\times${#2}}}

\newcommand{\RD}{\mbox{R\&D}\xspace}

\newcommand{\PhZ}{\mbox{Phase-0}\xspace}

\newcommand{\PhI}{\mbox{Phase-I}\xspace}

\newcommand{\PhII}{\mbox{Phase-II}\xspace}

\newcommand{\LumiUnit}{\mbox{cm$^{-2}$ s$^{-1}$}\xspace}

\newcommand{\PhIILumiMax}{\mbox{7$\times$10$^{34}$ cm$^{-2}$ s$^{-1}$}\xspace}

\title{ATLAS Upgrade Instrumentation in the US}

\usepackage{authblk}
\renewcommand\Authands{, }
\renewcommand\Affilfont{\itshape\small}

\author[a]{Gustaaf Brooijmans}
\author[b]{Hal Evans}
\author[c]{Abe Seiden}

\affil[a]{Columbia University}
\affil[b]{Indiana University}
\affil[c]{University of California Santa Cruz}

\abstracttext{
  Planned upgrades of the LHC over the next decade should allow the
  machine to operate at a center of mass energy of 14 TeV with
  instantaneous luminosities in the range 
  5--\PhIILumiMax .
  With these parameters, ATLAS could collect
  3,000 \invfb\ of data in approximately 10 years.
  However, the conditions under which this data would be acquired are
  much harsher than those currently encountered at the LHC. For
  example, the number of proton-proton interactions per bunch crossing
  will rise from the level of 20--30 per 50 ns crossing observed in
  2012 to 140--200 every 25 ns.
  In order to deepen our understanding of the newly discovered Higgs
  boson and to extend our searches for physics beyond that new
  particle, the ATLAS detector, trigger, and readout will have
  to undergo significant upgrades. In this whitepaper we describe
  \RD\ necessary for ATLAS to continue to run effectively at the
  highest luminosities foreseen from the LHC. Emphasis is placed on
  those \RD\ efforts in which US institutions are playing a leading
  role. 
}

\begin{document}

\tableofcontents
\clearpage

\section{Introduction}
\label{sect:intro}
The LHC is expected to be the world's Energy Frontier machine for
at least the next 15 years.  Data to be collected will allow much more
detailed studies of the Higgs boson, the search for other new
phenomena and the study of these if found, and in general push the
sensitivity for new physics well into the multi-TeV
range~\cite{SnowmassPhysWP}. This
physics program provides the challenge of simultaneously looking at
many decay and production channels for a 125 GeV particle, resulting
in final state particles of moderate tranverse momenta, and also
searching for particles at the highest mass scale that can be probed
by the accelerator, which can result in very high momentum particles
and large missing energy, all with very many superposed pile-up
events.

The mass of the Higgs boson is ideal for the study of many decay and
production channels, allowing an incisive test of our model for
electro-weak symmetry breaking.  However, present data samples are
quite limited and all channels are still crudely measured. Providing
much more precise measurements will require much more data.  A major
goal of the \PhII\ upgrade of the LHC is to provide approximately
300 times as many events in the various Higgs channels as presently
exist (and 10 times what will exist prior to the upgrade).  This
sample will also open up some new channels with small rates to provide
a more comprehensive Higgs picture.  In addition, to more fully
complete the picture of electro-weak symmetry breaking, the
measurement of W-W scattering at the TeV mass scale is very
important.  Acquiring such an event sample requires accelerator
upgrades in luminosity, and detector upgrades that maintain trigger
thresholds to enable collection of key events
while preserving a high quality detector with broad capabilities to
limit systematic errors on the quantities we wish to measure.

The presently planned machine upgrade will be to an instantaneous
luminosity of 5--\PhIILumiMax\ and would use luminosity leveling
at a 14 TeV center-of-mass energy to provide the largest number of
events. The goal is to collect 3,000 \invfb\ of data in approximately a
10 year time span.  The number of interactions will rise from the
present 20 to 30 per 50 nsec crossing to at least 140 and possibly 200
every 25 ns.

The \PhII\ detector will require fundamental upgrades in the trigger
and detector electronics across all systems to maintain thresholds and
efficiency, as well as an entirely new tracker to deal with the
increased radiation damage, occupancy, and data rates.  These
improvements will be based on advances in technology over the more
than 15 years since the present ATLAS detector was designed, and the
vigorous ATLAS upgrade \RD\ program now ongoing.  The various major
parts of the \PhII\ upgrade are described below with an emphasis on
US efforts.  A much more detailed discussion can be
found in the ATLAS \PhII\ Letter of Intent (LoI)~\cite{Ph2-LoI} 
with more background
material available in the \PhI\ LoI~\cite{Ph1-LoI}.

\graphicspath{{track/figures/}}

\section{Tracker Upgrade}
\label{sect:track}

\subsection{General Design Goals and R\&D Areas}
\label{sect:track:goals}
The design of the upgraded ATLAS tracker is governed by the physics
needs of the experiment and the desire to minimize cost.  The design
under consideration is based entirely on layers of pixel and strip
silicon sensors, which can provide the granularity and radiation
hardness required for the tracker.  Minimizing cost and material in
the tracking volume requires the fewest layers that can provide robust
pattern recognition and track parameter measurement.  This has been
determined by simulations to be four pixel layers surrounded by five
double-sided strip layers spread out over the 1 meter radius tracking
volume inside the ATLAS solenoidal magnet. Maintaining an occupancy
below about 1\% per sensor element allows good matching of strip
layers arranged to provide small angle stereo information with few
fake tracks. In order to keep the occupancy below 1\%, we must reduce
each cell size, resulting in an increase in the number of readout
channels by an order of magnitude compared to the present ATLAS
detector.  The inner radius of the less costly strip layers is
determined by radiation damage and the desire to have at least half
the signal strength from a passing particle, as compared to an
unirradiated detector, after collecting 3,000 \invfb\ of data.

To identify many of the physics signatures, the tracker must provide
excellent momentum measurements for energetic leptons and establish
that they are in fact isolated.  The material in the tracker affects
the efficiency and resolution in the detection of isolated electrons, due to
bremsstrahlung and nuclear interactions. The performance of $b$-tagging
algorithms based on displaced vertices also requires controlling the
tails in the impact parameter measurement due to errors in pattern
recognition and hadronic interactions within the active volume of the
tracker.  The upgraded tracker therefore must maintain careful
control of the material inside the tracking volume, despite the larger
channel count. This has significant implications on the mechanical
design of the support structure, the powering scheme, and the cooling.
In addition new physics analysis techniques have emerged that place
stringent requirements on future trackers.  For example, the use of
highly boosted jets require that track reconstruction work efficiently
in very collimated jets, where particle tracks remain close to each
other over long distances.  This demands small pixels and thin sensors
to avoid large cluster sizes.

An important lesson from the 8 TeV run is that tracking can be used to
build powerful tools to mitigate the effects of pileup for a wide
range of objects measured by other subdetectors. Therefore, another
important requirement on the upgraded tracker is the need to provide
fast momentum information on tracks to select events of interest at
the first trigger level.  Since the readout and reconstruction of all
tracks for a first level trigger is presently beyond the
state-of-the-art, the ATLAS plan is to only readout a fraction of the
tracking detector ($\sim$5\%) for participation in the first level trigger,
chosen via regions of interest based on information from the
calorimeter and muon systems.
This requires high-bandwidth frontend and controller
chips that allow for readout paths both for triggering and more
general readout.

The large integrated fluences for \PhII\ will require more radiation
hard sensors than in the present ATLAS detector but also at a minimum
cost.  In the case of pixels the goal has been to develop larger
detectors (for example 4cm $\times$ 4cm in area) that can be bump bonded to
several (for example four) frontend chips, significantly reducing bump
bonding costs.  For the strip sensors, units that are about 10cm $\times$
10cm allow cost minimization with further segmentation provided by
reading out shorter strip segments (for example 2.5cm length) using
hybrids and electronics located on top of the sensor.  Maintaining the
sensor signal after large fluences requires the collection of
electrons.  For most of the pixels and all of the strips cost
minimization then determines that the sensor be of the n-on-p type.
For the two very inner pixel layers, 3-D sensors, where readout
collection columns are embedded in the silicon, allow very short
collection distances and the highest radiation tolerance. The use of
slim-edges for the sensors will avoid having to overlap them to
maintain hermetic coverage, reducing material and the number of sensor
units.

The issues described above lead to a number of major \RD\ areas.
US groups have played a large role in many of the analogous areas
for the present ATLAS detector. Currently, three national lab groups
and six university groups are playing a major role in the
\RD\ program for \PhII .  Some primary areas in that program are:
\begin{itemize}
  \item Development of minimum cost sensor assemblies in n-on-p
    technology, aimed at both the strip detector and outer pixel
    layers.
  \item Optimizing 3-D sensors for the two inner pixel layers.
    Other options for this part of the detector are diamond sensors
    or planar detectors that can run at rather high voltages.  In
    addition new approaches, for example CMOS sensors, could have a
    large impact but require significantly more development and
    testing. 
  \item Developing slim-edge sensors.
  \item Developing pixel frontend chips, with those for the inner two
    layers being most challenging and requiring very high performance
    CMOS technology. 
  \item Development of strip front-end electronics, including features
    for first level triggering. 
  \item Developing very light-weight supports for mass minimization.
  \item Developing novel powering schemes to save on cabling, mass,
    and cost.
  \item Developing chips which provide for interfaces to the data
    acquisition and also to the powering system.
  \item Developing fabrication schemes that minimize assembly time and
    complexity.
  \item Developing cooling systems with minimum mass, which are robust
    and allow running well below room temperature.
\end{itemize}

An additional area of importance is the optical fiber system, the
optical transmission components, and read-out-drivers.  This is an
area where options from industry develop rapidly over time and we are
mainly following options available commercially.  However, high
bandwidth readout is expected to be of great importance in moving the
information from the sensors to a region at larger radius.
 
All of the newly developed items must be tested for radiation hardness
(e.g.~signal properties for sensors; speed, gain changes, and single
event upset for electronics), which leads to a very demanding and
time-consuming test program.

The next section presents the performance of the tracker we expect to
be able to build given several more years of \RD .  This assumes that
the \RD\ projects listed above converge to “construction-ready” status
over that time period.  The \RD\ over the past few years and
performance of the present detector provide some confidence that this
can be achieved given adequate funding.

\subsection{ATLAS Letter of Intent Tracker}
\label{sect:track:loi}
\begin{figure}[htbp]
\begin{center}
  \includegraphics[width=0.75\textwidth]{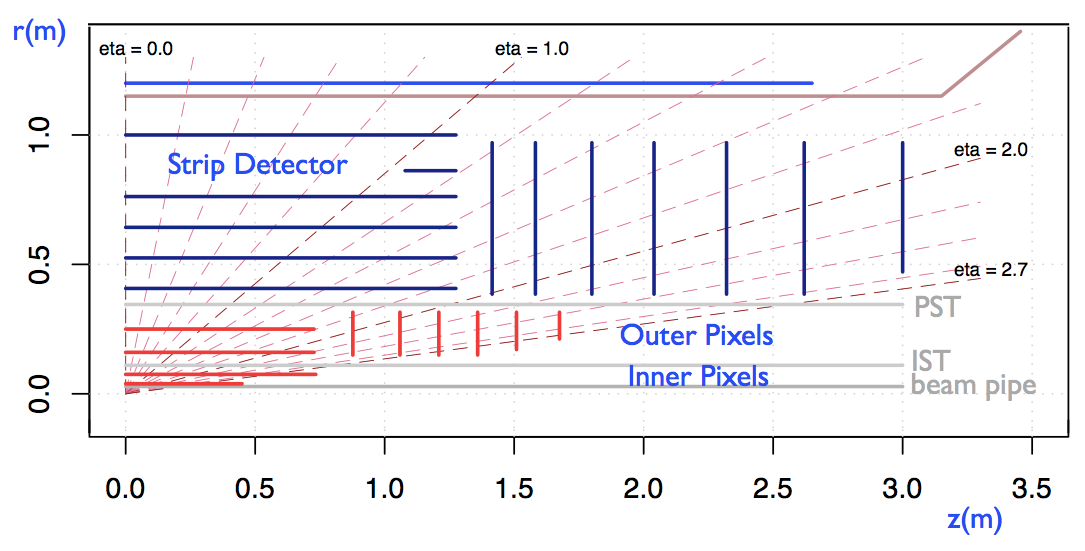}
  \caption{
    Diagram showing the layout of the tracker presented by ATLAS in
    the \PhII\ LoI.
  }
  \label{fig:track:layout}
\end{center}
\end{figure}

\begin{figure}[htbp]
\begin{center}
  \subfigure[]{
    \label{fig:track:hits}
    \includegraphics[width=0.450\textwidth]{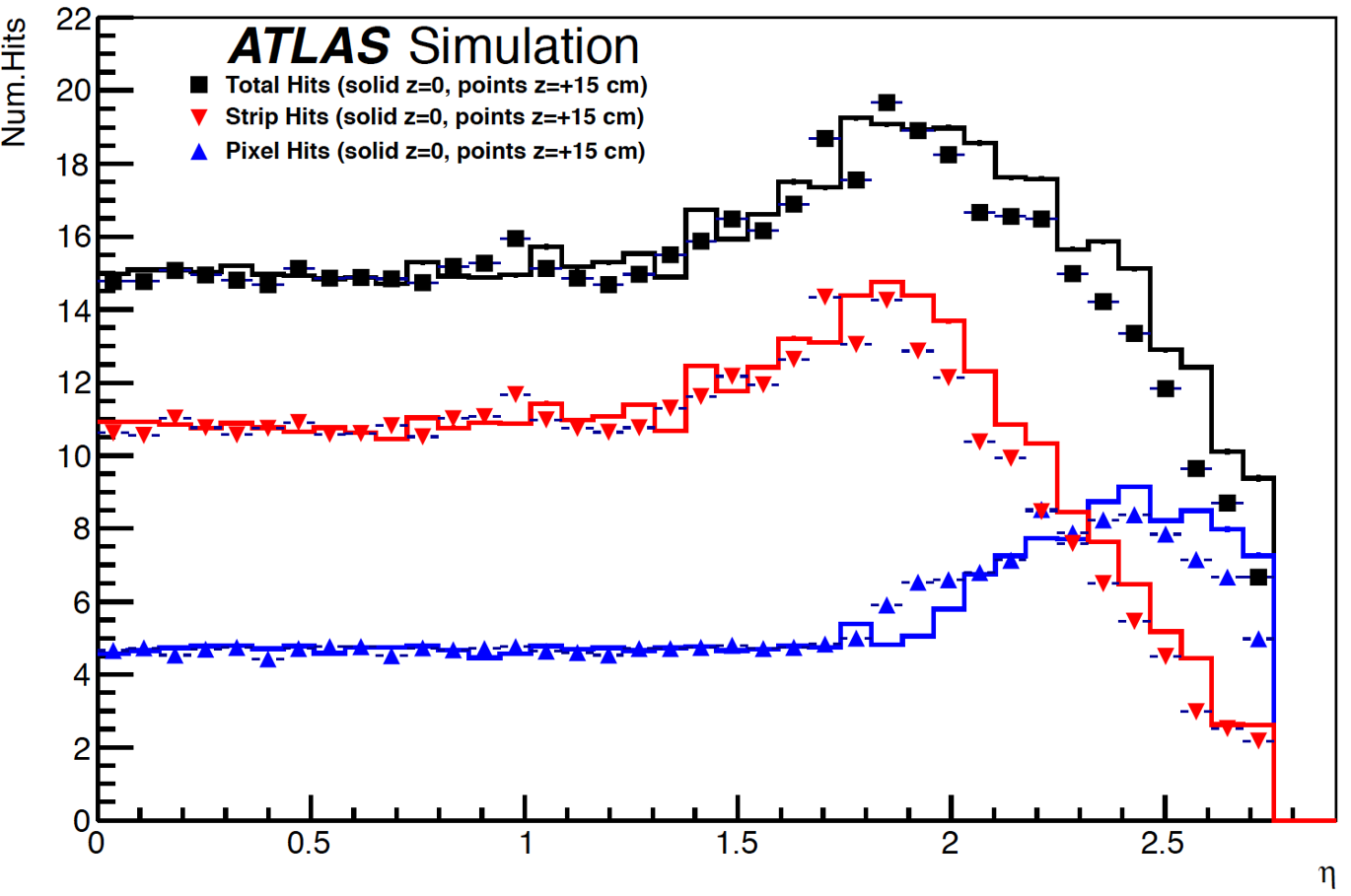}}
  \subfigure[]{
    \label{fig:track:material}
    \includegraphics[width=0.450\textwidth]{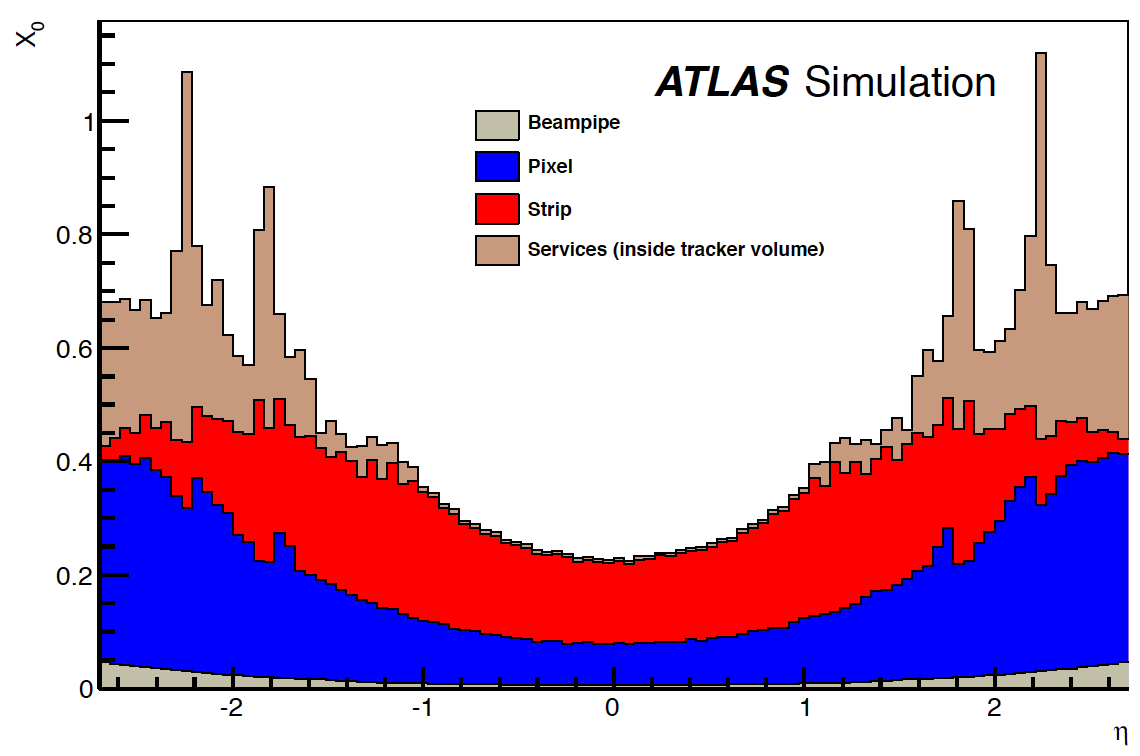}}
  \caption{
    \subref{fig:track:hits} Number of hits on a high momentum track
    originating from the origin (solid) and 15cm from interaction point
    (points), broken down between pixel hits, hits on strip detector,
    and total hits.
    \subref{fig:track:material} The material traversed by tracks from
    the origin, broken down by source of material.
  }
  \label{fig:track:coverage}
\end{center}
\end{figure}

\begin{figure}[htbp]
\begin{center}
  \includegraphics[width=0.50\textwidth]{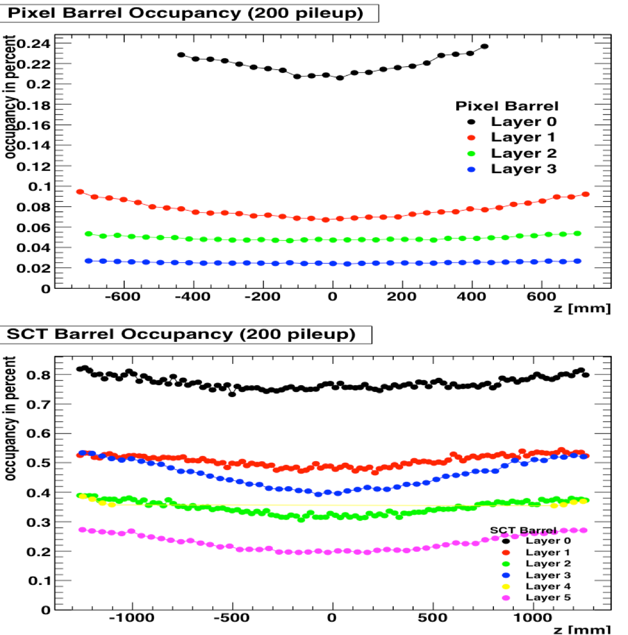}
  \caption{
    Occupancy expected in the LoI tracker for 200 interactions per
    crossing.
  }
  \label{fig:track:occ}
\end{center}
\end{figure}

Figure~\ref{fig:track:layout} shows the tracker that has been
simulated and on which the 
following performance plots are based.  It consists of four pixel
layers and five double-sided strip layers in a long barrel, and a
collection of disk layers to cover more forward rapidities.  The use
of a barrel that is much longer than in the present ATLAS detector
leads to a reduction in material as well as pushing material to larger
rapidities.  The layout has been adjusted to maintain 14 precision
measurements on nearly all tracks, required for good pattern
recognition.  
Figure~\ref{fig:track:coverage}~\subref{fig:track:hits}
shows the number of hits on tracks originating
from the origin and also from 15cm along the beam-line, indicating the
excellent coverage.
Figure~\ref{fig:track:coverage}~\subref{fig:track:material} 
shows the material traversed (in
radiation lengths) by a high 
momentum track as a function of rapidity.  The different sources of
material are indicated.  The layout arranges for a minimum amount of
material at rapidity below 1.6, with most of the services beyond this.
Figure~\ref{fig:track:occ} 
shows the occupancy expected in the tracker
for each barrel layer for the case of 200 interactions in a crossing.

\subsection{Performance Expectations}
\label{sect:track:performance}
\begin{figure}[htbp]
\begin{center}
  \subfigure[]{
    \label{fig:track:pTres}
    \includegraphics[width=0.60\textwidth]{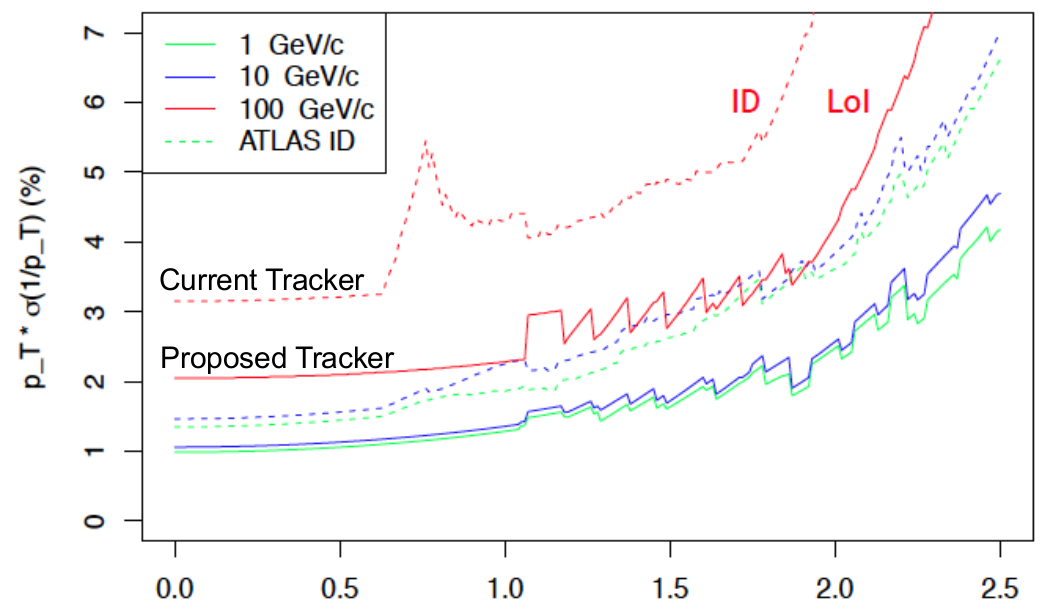}}
  \subfigure[]{
    \label{fig:track:ratio}
    \includegraphics[width=0.45\textwidth]{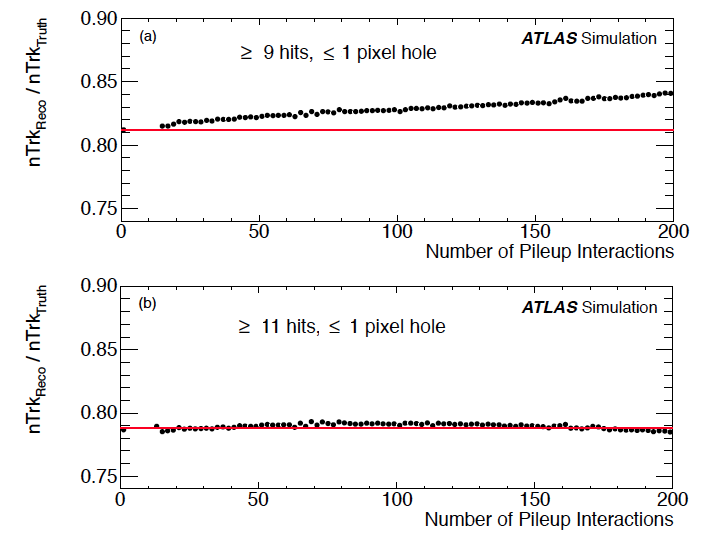}}
  \subfigure[]{
    \label{fig:track:b-tag}
    \includegraphics[width=0.45\textwidth]{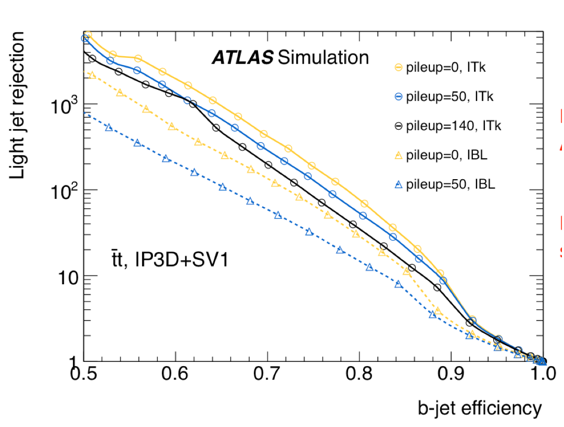}}
  \caption{
    \subref{fig:track:pTres} Predicted momentum resolution for muons
    as a function of 
    pseudorapidity for the new inner tracker (solid lines) compared
    with the current tracker (dotted lines) for three different values
    of the momentum.
    \subref{fig:track:ratio} Ratio of reconstructed to generated
    tracks from simulated 
    $t\bar{t}$ events at various levels of pileup for two different
    track selections: requiring track reconstruction with at least 9
    hits per track (b-a), and with at least 11 hits per track (b-b).
    \subref{fig:track:b-tag} Expected b-tagging performance, the
    efficiency to tag a $b$-jet 
    and related rejection factor for light-quark jets.
  }
  \label{fig:track:perf}
\end{center}
\end{figure}

The tracker is designed for physics at up to 200 events per crossing.
We present below some of the performance expectations.   
Figure~\ref{fig:track:perf}~\subref{fig:track:pTres}
shows the expected momentum resolution for muons of three different
transverse momenta.  Despite the discrete layer spacing the tracker
achieves a rather smooth distribution versus rapidity and improves on
the present tracker.
Figure~\ref{fig:track:perf}~\subref{fig:track:ratio} 
shows the ratio of reconstructed to generated tracks from
$t\bar{t}$ events as a function of pileup for two different track
selections: requiring track reconstruction with at least 9 hits per
track (a), and with at least 11 hits per track (b).  The choice of 14
hits allows us to achieve good performance (that is no increase in
fake tracks with pileup) with some redundancy, for example with some
missing hits (1 pixel layer and 1 double-sided strip layer).
Figure~\ref{fig:track:perf}~\subref{fig:track:b-tag}
shows the expected $b$-tagging performance for different levels
of pileup as well as a comparison to the expectations for the present
tracker, including the new Insertable B-Layer (IBL).  The performance
for b-tagging of the upgraded tracker (labeled ITK) with 140 events
per crossing is approximately as good as that for the present tracker
with no extra occupancy.  This level of performance is also crucial
for reconstructing $\tau$ leptons.

\graphicspath{{calo/figures/}}

\section{Calorimetry}
\label{sect:calo}
Calorimetry at high-luminosity, high-energy hadron colliders faces a
number of major challenges:
\begin{itemize}
  \item Front-end electronics need to be radiation-tolerant and
    consume little power while achieving high-precision (typically
    16-bit dynamic range) and high data transmission bandwidth.
    Furthermore, these usually need to be installed on-detector,
    i.e.~in inaccessible locations that impose stringent reliability
    requirements.
  \item High radiation levels and particle fluxes, particularly in the
    forward regions, impose the use of technologies that are
    simultaneously very radiation tolerant and able to yield
    high-precision measurements in the presence of high background
    fluxes.
\end{itemize}

The value of high-precision calorimetry to accurately measure
electrons, photons and hadronic jets has been well established by the
discovery of the Higgs boson and e.g.~precision measurements in top
quark physics~\cite{SnowmassPhysWP}.
During \PhII\ operation, particle fluxes and the
average energy deposited in the calorimeters are expected to be
typically five to ten times higher than specified in the LHC design
values. Under these conditions it is certain that the front-end
electronics will have to be replaced. This is due both to radiation
damage and the need for ATLAS to upgrade the trigger system, with the
latter requiring real-time performance capabilities that the current
electronics cannot satisfy.

The ATLAS calorimeters are based on two technologies: liquid argon
sampling for the electromagnetic and forward hadronic calorimeters,
and scintillating tile sampling for the central hadronic calorimeter.
In both cases, the front-end electronics currently send coarse data to
the Level-1 trigger system while buffering the precision data for
readout on Level-1 accept.  Ongoing \RD\ work is aimed at exploiting
technological progress to send the full precision data to the Level-1
trigger system for all beam bunch crossings.

The US has played a leading role in the development and construction of the 
initial ATLAS calorimeter electronics: most of the ASICs for the
liquid argon front-end electronics were developed in the US, and the 1524 
front-end boards were produced in the US.  The liquid argon back-end 
electronics were initially designed in the US.  For the TileCal, the US 
led the front-end electronics design and played a major role in their 
construction.

\subsection{Liquid Argon Front-End Electronics}
\label{sect:calo:fe-lar}
\begin{figure}[htbp]
\begin{center}
  \includegraphics[width=0.75\textwidth]{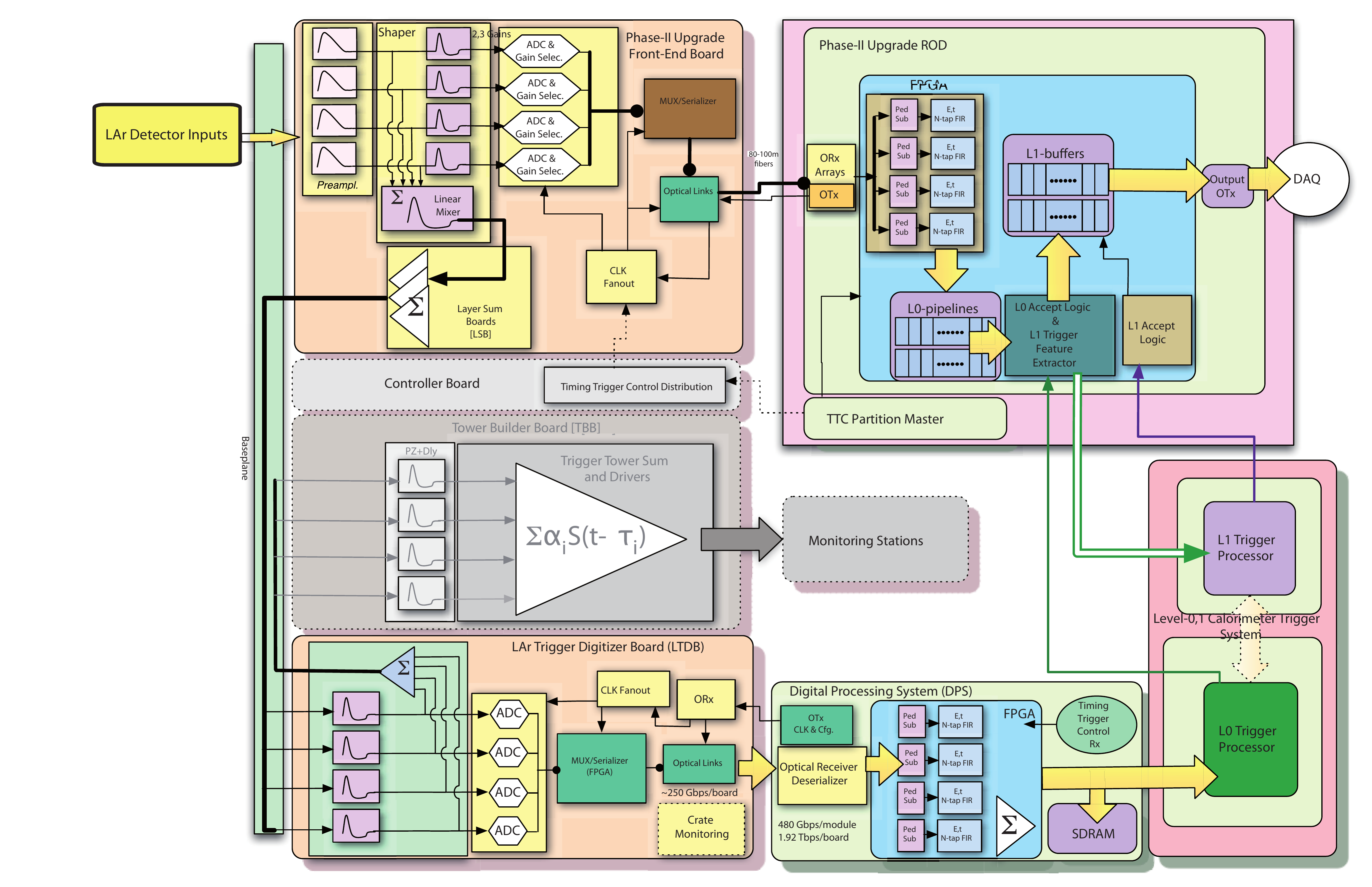}
  \caption{
    \PhII\ readout architecture for the liquid argon calorimeters.
    The front-end and LAr Trigger Digitizer boards are located
    on-detector, and send the data over fast optical links to the
    off-detector ROD and Digital Processing System.  The LTDB and DPS
    will be installed in the \PhI\ upgrades in 2018.
  }
  \label{fig:calo:PhII}
\end{center}
\end{figure}

The main architectural difference between the existing readout system
and the planned upgrade is the switch from an (analog) on-detector
Level-1 pipeline to a “free-running” design in which signals from all
calorimeter cells are digitized at 40 MHz and sent off-detector.  This
approach effectively removes all constraints imposed by the liquid
argon calorimeter (LAr) readout on the trigger system, since full
precision, full granularity data will be available and the latency and
Level-1 bandwidth will become essentially unlimited.  Furthermore, due
to its minimal coupling to the trigger system, the new architecture
will provide significant flexibility for further evolutions in the
ATLAS trigger system or overall detector readout.  It will be able to
accommodate a trigger system with a low-latency, low-granularity
Level-0 trigger stage just as well as a system with a high-latency,
full-detector Level-1 trigger, or both.
Figure~\ref{fig:calo:PhII} shows the readout architecture for \PhII .

The front-end boards (FEB) perform analog shaping of the calorimeter
signals, then sample the analog waveform at 40 MHz before multiplexing
the digitized outputs for transmission over fast optical links. (The
LAr trigger digitizer boards perform very similar tasks with somewhat
worse precision and coarser granularity to provide a more manageable
data volume for the Level-0 trigger.)  Each FEB reads out 128
calorimeter channels and transmits data at approximately 100 Gbps.
A total of 1524 FEBs are needed to read out the full calorimeter.

The new front-end board architecture is designed to remove the
bottlenecks inherent in the current boards by exploiting progress in
technology, while maintaining the analog performance: 16-bit dynamic
range (currently achieved with three gains at 12 bits each) and
coherent noise below 5\% of the incoherent noise level.  The main
functional blocks are: analog pre-amplification and shaping,
production of (summed) analog signals for the LTDB, analog-to-digital
conversion of all signals at a rate of 40 MHz, multiplexing and
serialization of digital data, and transmission over high-speed
optical links.  Since there will be no Level-1 pipeline, the control
logic will be limited to clock distribution and slow controls for
configuration and calibration.

US \RD\ Efforts address all aspects of the necessary front-end
electronics development.  The analog preamplification and shaping
stages will be integrated in a single ASIC.  The Liquid Argon Preamp
and Shaper (LAPAS) test-chip, designed by the University of
Pennsylvania and Brookhaven National Laboratory, and fabricated in
2009 in IBM's 8WL SiGe process, validated the design approach of
implementing a wide dynamic range single ended preamp followed by low
power differential shaping stages with multiple gains to achieve the
required 16 bit resolution.  An integral non-linearity of less than
0.6\% was demonstrated in both the 1$\times$ and 10$\times$
gain shaping stages and
measurements of ionizing radiation indicated robust performance for
TID in excess of 1 MRad~\cite{dressnandt2009a}.
Less expensive SiGe process alternatives
are currently being explored: IHP SG25H3P and IBM's 7WL.

The requirements on the ADC are a large dynamic range (16 bits can be
achieved using three gains and 12-bit ADCs if needed) and precision
(at least 12 bits), low power ($<$ 100 mW/channel), high speed (40
MSPS), small footprint (128 channels on a 50 cm-wide board) imposing
serialized outputs, and substantial radiation tolerance.  Many
commercial devices meet most of these criteria but are very sensitive
to single event upsets (SEUs). Irradiation tests are being performed
to verify if recent devices are suitable.

\begin{figure}[htbp]
\begin{center}
  \includegraphics[width=0.75\textwidth]{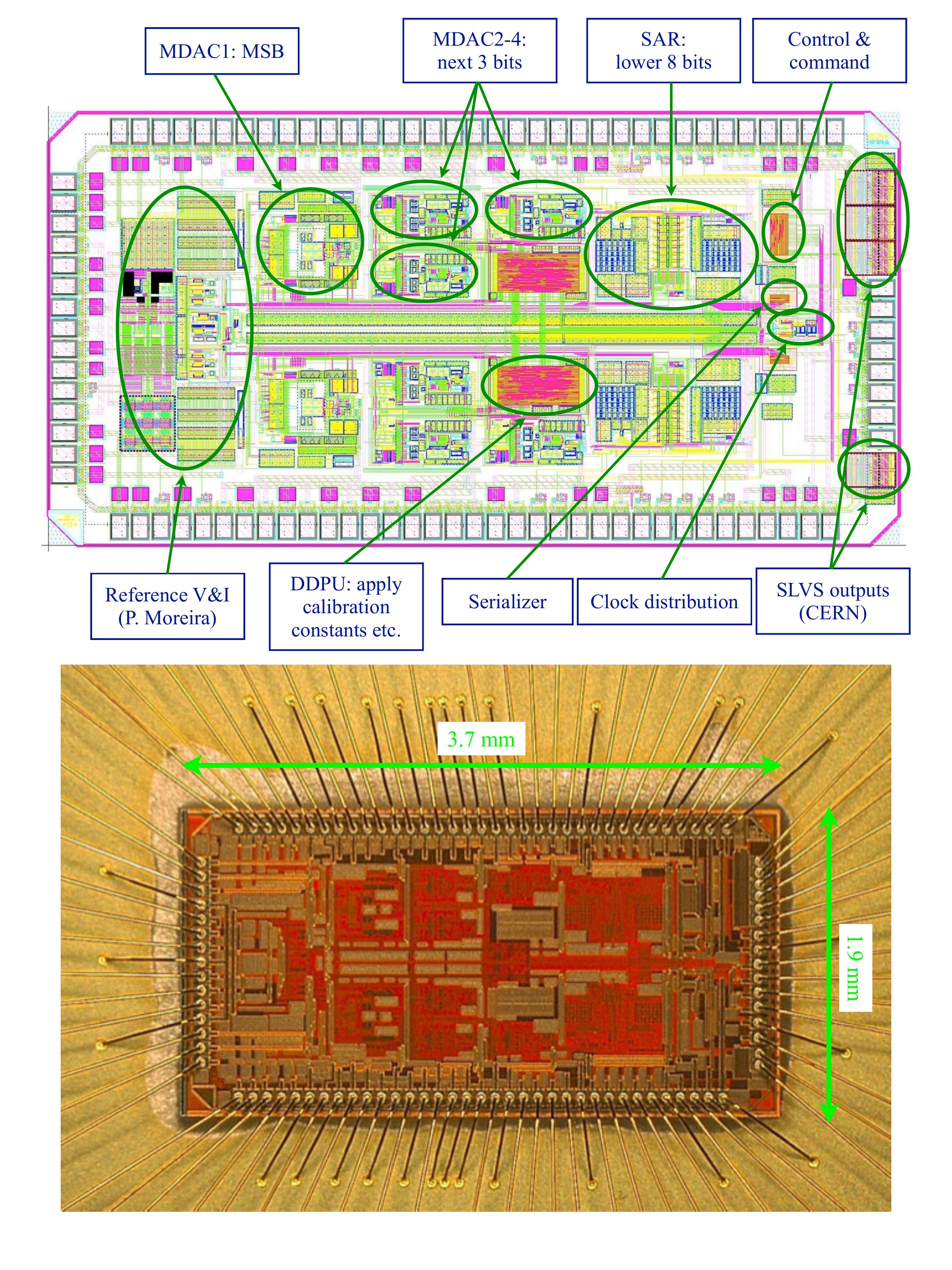}
  \caption{
    Schematic diagram and photo of the nevis12 chip: a dual channel,
    12-bit ADC test chip.
  }
  \label{fig:calo:nevis12}
\end{center}
\end{figure}

In addition to radiation tolerance, ASIC solutions have lower latency
and power consumption.  The former aspect is crucial if the signals
are to be used in the Level-0 or Level-1 trigger systems.  A recent
test chip (nevis12, developed at Columbia University’s Nevis Labs) in
IBM CMOS 8RF (130 nm) technology uses four 1.5-bit pipeline stages
followed by an 8-bit SAR.  The output data are serialized at 640 Mbps.
A predecessor has allowed to demonstrate analog performance of the
pipeline part at least equivalent to that of a commercial 12-bit ADC
together with excellent radiation tolerance, and early test results
from nevis12 are promising.  Full testing will be completed in 2013 to
confirm its analog precision, radiation tolerance, and determine the
optimal operational parameters.  The power consumption is expected to
be approximately 40 mW/channel (40\% of typical COTS ADCs) and the
latency 65 ns (100 ns less than the typical COTS pipeline ADC).
Figure~\ref{fig:calo:nevis12} shows a schematic diagram of the nevis12 chip.

After digitization, the raw data produced on a FEB represents 128
(channels) $\times$ 14 bits (including two gain scale bits)
$\times$ approximately
40 MHz, or about 72 Gbps. Conservatively assuming 25\% overhead for
control words and encoding, each board needs to multiplex, serialize
and transmit 90 Gbps of data, leading to a total data transmission
rate of 140 Tbps for the full calorimeter.  A radiation-tolerant
serializer in 250 nm silicon-on-sapphire technology able to run at 8
Gbps has already been developed by Southern Methodist University, and
with a newer process, speeds of 10 Gbps should be achievable.  It
should therefore be possible to use a single 12-fiber ribbon to
transmit the data produced by each FEB.

Further US ASIC \RD\ plans include: the production of a quad-channel
12-bit ADC, a 5.5 Gbps multiplexer and serializer chip and a 5.5 Gbps
laser driver chip for the \PhI\ upgrades; developing a 16-bit
dynamic range (12-bit precision) ADC which could possibly incorporate
the pre-amplification and shaping stages; incorporation of the ADC and
high speed (5-10 Gbps) output serializers.

An oft-overlooked but crucial issue is the power distribution to the
front-end electronics.  The powering concept for \PhII\ is based on a
distributed power architecture with crate-level main converters
providing 48 V and board-level point-of-load converters (POL). Modular
main converters with 1.5 kW per module are under development with
conversion efficiencies above 80\%. Operation in external fields up to
300 Gauss, as present at the location on the LAr detector, has been
successful.  POL converters need to operate in even stronger magnetic
fields, which prevents the usage of magnetic materials for inductance
cores.  Several developments of air-core based POLs are currently
ongoing at Brookhaven National Laboratory and different technologies,
based on SiCMOS and GaN, are under study.

\subsection{Tile Calorimeter Front-End Electronics}
\label{sect:calo:fe-tile}
The goals of the Tile Calorimeter \PhII\ upgrade are to increase the
radiation tolerance of the front-end electronics and to maximize the
resolution and configurability of the tower triggers by sending all
tile cell data to the sROD off-detector. This is feasible with the
advent of $>$ 10 Gbps optical communications.

\begin{figure}[htbp]
\begin{center}
  \includegraphics[width=0.75\textwidth]{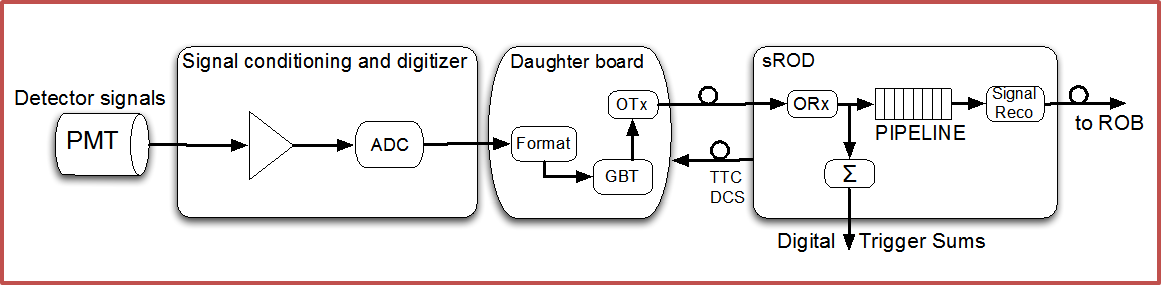}
  \caption{
    Tile electronics in \PhII .
  }
  \label{fig:calo:tile}
\end{center}
\end{figure}

Achieving these goals requires replacement of all the on-detector
electronics as well as the sROD off-detector readout.  Preservation of
data integrity is achieved in three stages.  The first line of defense
is achieved by using better radiation tolerant components.  Next,
there is the creation of redundancy in the on-board electronics so
that data is preserved if one power supply or microprocessor
fails. Additionally, tile cells are read out by two phototubes sent to
separate readout channels. Finally, error correction is employed in
the serial communication scheme.  An overview of the tile electronics
system is shown in the Figure~\ref{fig:calo:tile}.

In the upgraded scheme, the on-detector electronics are housed in a
mechanical drawer on which are mounted phototubes, an
amplifier/shaper card (which might also digitize the signal), a Main
Board which distributes low voltage and control signals, holds the ADCs,  and
routes data to a high-speed Daughter Board where the error correction
and communications processing occurs. Additionally, the drawer houses
a High Voltage control card for the phototubes. The Main Board is a complex
69 cm PCB of 14 layers, currently prototyped by the University of Chicago group.
This board must handle all controls, signal digitization, and timing delays.

Integrity of the Low- and High-Voltage systems is paramount to the reliability of the
Tile system, and both have been a source of problems in the past. 
The Argonne group has undertaken to redesign both systems. Their earlier 
redesign of the low-voltage modules for the current detector reduced noise and
increased radiation tolerance by an order of magnitude.

The Daughter Board preserves the philosophy of redundancy, with each
half handling 6 phototubes with a separate Kintex-7 FPGA. The DB
handles the high-speed optical communications, which is currently
envisioned to occur using radiation tolerant 40 Gbps optical
modulators.  This communications pipeline will send all tile cell data
to the sROD at 40 MHz for recording and formation of a digital
trigger. An additional benefit of the optical communication of the
trigger (as opposed to the current analog tower sum sent over 70 m
cables) is much lower noise, allowing for better jet trigger turn-on
resolution.

The Argonne and Chicago groups are also conducting \RD\ on optical communications.
At ANL, modifications of a Luxtera modulator-based optical transmitter are being studied
and prototypes will be radiation tested. Modulators are faster than commercial vcsel-based transmitters,
achieving transmission rates of greater than 10 Gbps with very low error rates and likely high radiation
tolerance. Use of a modulator rather than a vcsel transmitter could reduce system cost considerably.
The University of Chicago group is setting up a radiation test facility at FNAL to study the
single event upset rate in the optical transmitters and FPGA controls. 
This test facility will simulate the HL-LHC cavern environment.

Currently \RD\ is being conducted on three potential
front-end amplifier/shaper boards.  These boards process the phototube
signal and send it to the Main Board, control the calibrations via
charge injection and also the calibration achieved using a Cs source
(this requires a slow integrator).  A redesign of the current “3-in-1”
card has been prototyped by the University of Chicago group and has been 
shown to be radiation tolerant and
extremely linear; this card sends LVDS analog data to ADCs on the Main
Board.

Two alternative front end boards are also being studied. These are
ASIC-based with internal ADCs.  One version is a joint effort of the 
Argonne ATLAS group and the FNAL CMS group
to produce a new version of the QIE ASIC. This chip integrates the
input current and digitizes in 4 gain ranges. The other ASIC \RD\ uses
the current conveyor concept to digitize three gain ranges and
incorporates a custom radiation tolerant ADC.  The decision on which
front-end card to be incorporated in \PhII\ will only be made after
beam testing of all three options.

\subsection{Off-Detector Electronics}
\label{sect:calo:fe-off}
The main tasks of the off-detector electronics for the calorimeters
are the reception of digitized data from the FEBs, data filtering and
processing, as well as data transmission to the trigger and DAQ
systems.  The total input data rate for the LAr off-detector
electronics amounts to about 140 Tbps.

The central components will be Read-Out Driver boards (ROD)
implemented in ATCA standard.  These receive the data using serial
optical links on multi-fiber ribbons.  Conversion to electronic
signals is performed by commercial components and deserialization is
handled by fast FPGA transceivers.  The RODs will apply digital
filtering using FPGAs to calculate calibrated energy deposits together
with the signal time for each cell as well as signal quality
criteria. The digital filtering will correct for electronic and
pile-up noise and will be adjustable to the expected high-luminosity
conditions by parametrized FPGA algorithms. The RODs will buffer the
processed data in digital memory blocks according to the number of
hardware trigger levels and their corresponding trigger latencies,
until data are eventually transmitted to the DAQ system.  The data
from the individual LAr cells are further processed by the RODs in
order to provide input to the Level-1 hardware trigger. The signal
processing may go beyond simple sums of cell energy or transverse
energies. More complex algorithms like the extraction of features of
electromagnetic shower shapes or fast tagging of
$\pi^0 \to \gamma \gamma$ signals
using the finely segmented first calorimeter layer may be possible
given the availability of the full granularity of the LAr
calorimeters.  The detailed algorithmic layout is the subject of
on-going \RD , which will take into account how many channels can be
concentrated in one processing FPGA and the latency required by the
trigger system.

US \RD\ activities at Brookhaven National Laboratory, the State
University of New York at Stony Brook and the University of Arizona
tackle all aspects of this work: use of fast, small form-factor
optical transceivers, data handling and processing in high-end FPGAs,
and optimal exploitation of the ATCA standard for this application.
It should be noted that by going to ATCA, only six racks will be needed
to read out the full LAr calorimeter, to be compared with the 17 VME
crates currently installed, even though the data volume will have been
multiplied by more than 50.

For the TileCal, the groups at Michigan State University and the University of Texas at Arlington are central
to this effort.  They are working with engineers at CERN and IFIC Valencia to prototype the system
and devise detector control protocols.

\subsection{Summary of Main Calorimeter R\&D Areas}
\label{sect:calo:rd}

Areas in which US groups are playing a leading role are listed below.

\begin{enumerate}
\item {\bf Analog Signal Processing:}
 The liquid argon calorimeter signals need to be amplified and shaped to reduce sensitivity to 
 pile-up and electronics noise.  ASIC design efforts based on silicon-germanium bipolar technology
 have shown amplification and shaping can be integrated in a single chip without compromising 
 signal quality.
\item {\bf Analog to Digital Conversion:}
 The large number of channels and stringent specifications
 make ASIC ADC solutions very attractive and cost effective.
 Integration of additional, ATLAS-specific functionalities will allow to further
 optimize the readout architecture.
\item {\bf High Bandwidth Data Transmission:}
 The ATLAS calorimeters will transmit over 200 Tbps of data off-detector in \PhII.  Low power,
 radiation tolerant data transmission will be key to realize this on the front-end, while highly
 integrated solutions (optical links on FPGAs) would be very attractive off-detector.
\end{enumerate}

\graphicspath{{tdaq/figures/}}

\section{Trigger and Data Acquisition}
\label{sect:tdaq}
Running conditions anticipated during \PhII\ LHC operations cause
significant stresses on the ATLAS Trigger and Data Acquisition
(TDAQ) systems.
Luminosities of up to \PhIILumiMax , leading to an average number of
proton-proton collisions per bunch crossing, \aveMu , of approximately
200 every 25 ns, require major upgrades of the TDAQ in the areas of
dataflow and event processing capabilities. 
Work has already started in evaluating technologies that will allow us
to meet these challenges.
Many of these solutions are expected to take advantage of advances in
commercially available hardware yielding an upgraded system that is
more uniform and easily maintainable than the current TDAQ, but that
is also flexible enough to adjust to changes in our goals as
understanding of the physics driving the upgrades advances.

\subsection{Physics Motivation}
\label{sect:tdaq:phys}
Stresses on the TDAQ system
at luminosities expected to reach \PhIILumiMax\
are one of the main
motivations for \PhII\ upgrades of the ATLAS experiment. 
In order to achieve
the physics goals outlined in Ref.~\cite{SnowmassPhysWP}, the TDAQ will
have to maintain or improve upon performances achieved in 2012 running
and those foreseen to be realized in the ATLAS \PhI\ upgrade
program.
A brief summary of trigger requirements needed to maintain
sensitivity to important physics channels is given in
Table~\ref{table:tdaq:phys}. 

\begin{table}[htbp]
\begin{center}
  \caption{Trigger requirements to maintain sensitivity to selected
    physics channels.\label{table:tdaq:phys}}
\begin{tabular}{ccc}
\hline
  {\bf Channel} & {\bf Trigger} & {\bf Target Thresholds}\\
\hline
  $W,Z$ & single-lepton $e, \mu$ & \pT\ $\sim$ 20 GeV\\
  $t\bar{t}$ & di-jet & \ET\ $\sim$ 60--80 GeV\\
  $HH \to b \bar{b} \gamma \gamma$ & di-photon & \ET\ $\sim$ 10 GeV\\
  VBF & single-lepton + forward jets & $E_T^{\mathrm{jet}}$ $>$ 50 GeV\\
  SUSY & single-lepton + \ETmiss 
    & $p_T^{\ell}$ $\sim$ 20 GeV, \ETmiss\ $\sim$ 150 GeV\\
\hline
\end{tabular}
\end{center}
\end{table}

\begin{figure}[htbp]
\begin{center}
  \subfigure[]{
    \label{fig:tdaq:EMrate}
    \includegraphics[height=60mm]{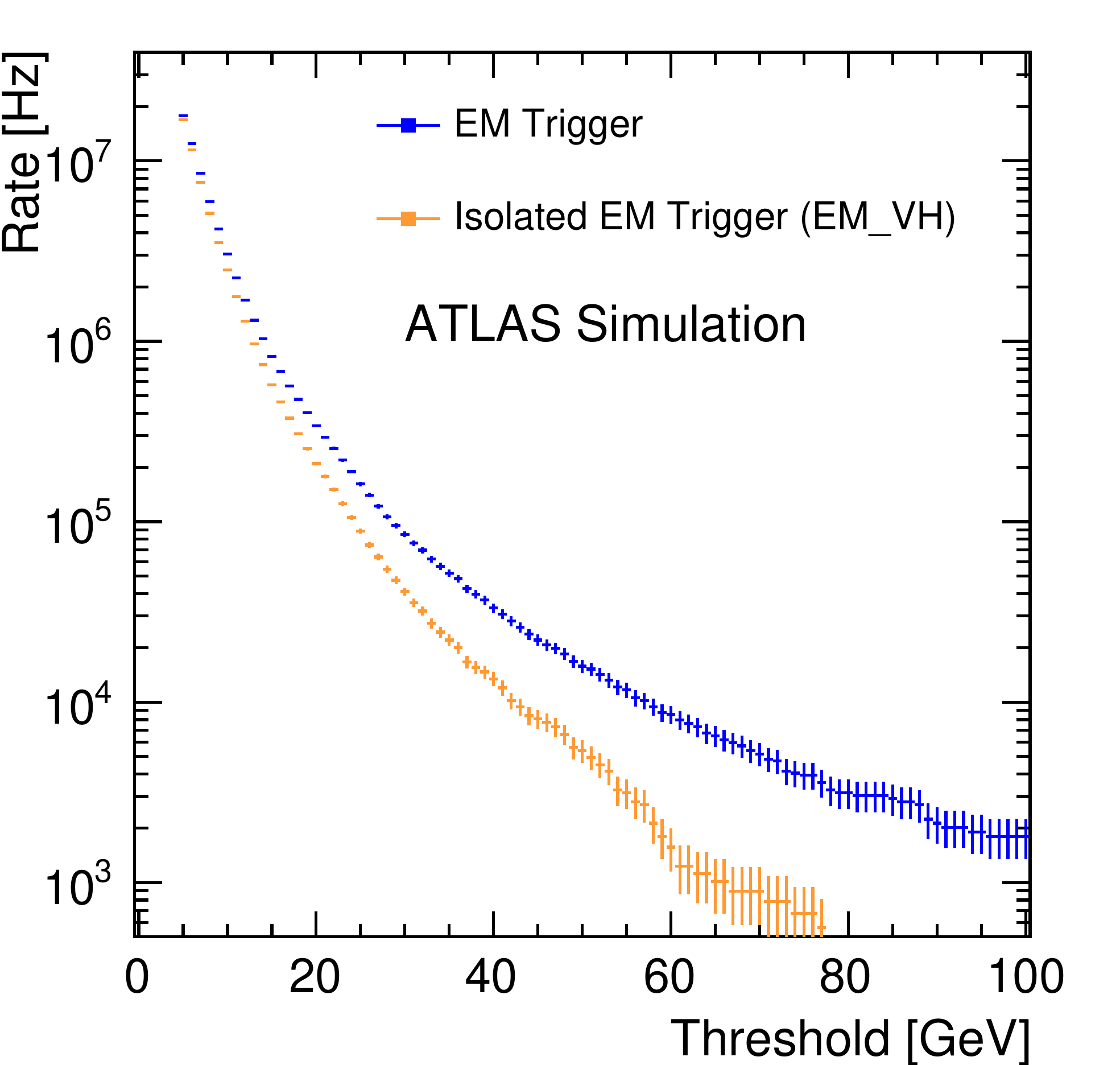}}
  \subfigure[]{
    \label{fig:tdaq:MuAccept}
    \includegraphics[height=60mm]{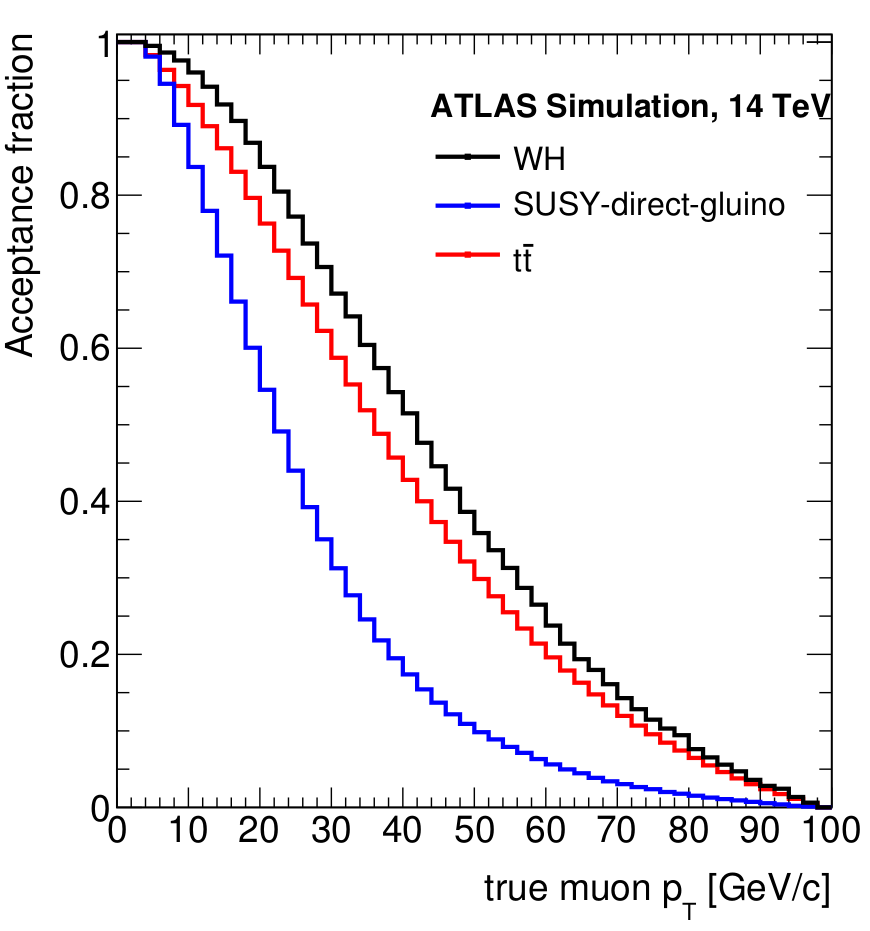}}
  \caption{
    \subref{fig:tdaq:EMrate} Projected EM trigger rates for
    the \PhI\ trigger system at
    \aveMu\ = 115, corresponding to 4$\times$10$^{34}$ \LumiUnit ,
    as a function of trigger threshold.
    \subref{fig:tdaq:MuAccept} Acceptance of muons from $t\bar{t}$,
    $WH$, and SUSY processes as a function of the true muon momentum.
  }
  \label{fig:tdaq:Leptons}
\end{center}
\end{figure}

All aspects of the ATLAS TDAQ system are impacted by these
requirements. However one of the main drivers of the \PhII\ TDAQ
upgrade is the desire to maintain Level-1 trigger thresholds for
isolated electrons and muons at around 20 GeV, with an accept rate of
20 kHz or less, for luminosities of up to
\PhIILumiMax .
Estimates of the Level-1 EM trigger rate vs.~threshold are shown in
Figure~\ref{fig:tdaq:Leptons}~\subref{fig:tdaq:EMrate} 
for the EM trigger configuration anticipated
in \PhI . Even for isolated EM triggers, a threshold of 40 GeV or more
would be required to achieve an accept rate of 20 kHz or
lower. However,
Figure~\ref{fig:tdaq:Leptons}~\subref{fig:tdaq:MuAccept} shows that
raising the Level-1 lepton trigger thresholds from their current
values (around 20 GeV) to the 40 GeV required to meet bandwidth
limitations would reduce acceptance by a factor of 1.5--3
depending on the physics channel.

\subsection{TDAQ System before \PhII }
\label{sect:tdaq:hist}
\begin{figure}[htbp]
\begin{center}
  \subfigure[]{
    \label{fig:tdaq:L1PhI}
    \includegraphics[width=0.50\textwidth]{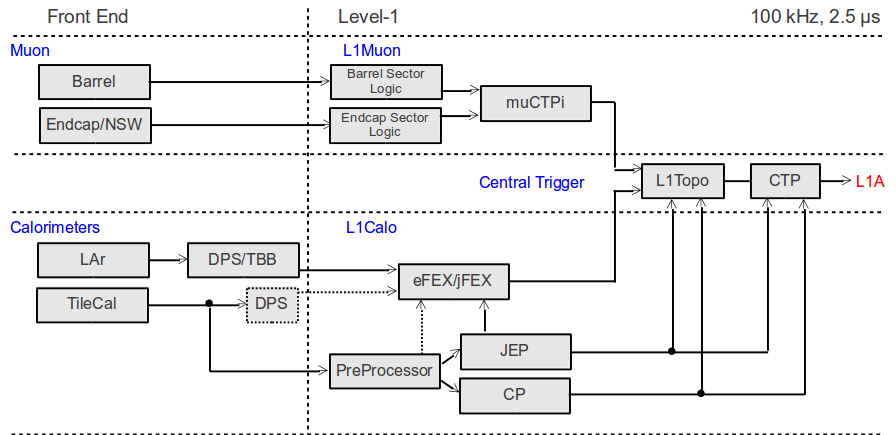}}
  \subfigure[]{
    \label{fig:tdaq:DaqPhI}
    \includegraphics[width=0.40\textwidth]{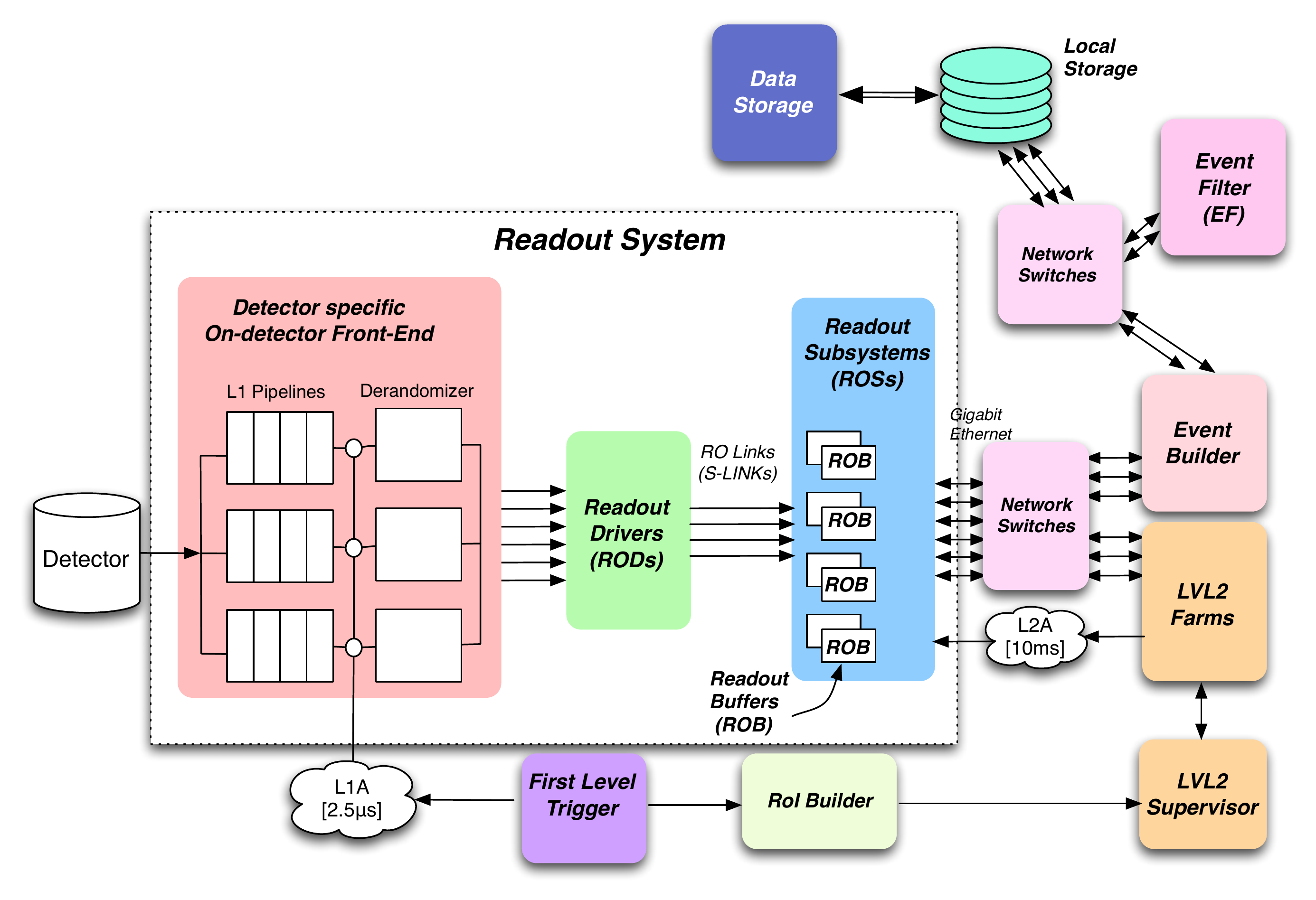}}
  \caption{
    Functional diagrams of:
    \subref{fig:tdaq:L1PhI} the L1 trigger system;
    and \subref{fig:tdaq:DaqPhI} the TDAQ
    in \PhI .
    Note that in \PhI\ EF, Event Builder, and Level-2 nodes will be
    implemented in the same processor.
  }
  \label{fig:tdaq:PhI}
\end{center}
\end{figure}

The trigger architecture currently used by ATLAS and foreseen through the
end of  \PhI\ running consists of three levels.
\begin{enumerate}
  \item {\bf Level-1 (L1)} uses
    custom hardware to construct trigger {\it Regions of Interest}
    (ROIs)  based on information from the
    Calorimeters and Muon system
    and takes advantage of correlations between objects using
    a Topological Trigger Processor.
    During \PhI\ running the Level-1
    trigger system will generate trigger accepts at an average rate of
    100 kHz, with a latency of approximately 2.5 \us .
    An overview of the L1 system during \PhI\ operation is given in
    Figure~\ref{fig:tdaq:PhI}~\subref{fig:tdaq:L1PhI}. 
  \item {\bf Level-2 (L2)} is implemented in software on a farm of PCs and
    constructs triggers using information from all ATLAS sub-systems,
    but only in  ROIs identified by L1.
  \item {\bf Event Filter (EF)} runs on the same farm of PCs as L2, but has
    access to data from the full event and uses algorithms similar, if
    not identical, to those run in offline reconstruction. Trigger
    accepts from the EF at a rate of up to 1 kHz are foreseen in \PhI .
\end{enumerate}
L2 and the EF, which in \PhI\ will run in the
same CPU along with Event Building, 
are collectively referred to as the {\it High Level Trigger} (HLT). 

The ATLAS DAQ is centered around system-specific Readout Drivers
(RODs) that collect data from the individual sub-system front ends
(FE) on L1 accepts, and the common Readout System (ROSs) that
provides this data on request to the HLT.
The configuration of the DAQ system, as it is foreseen in \PhI\
running, is shown in 
Figure~\ref{fig:tdaq:PhI}~\subref{fig:tdaq:DaqPhI}. 

\subsubsection{Involvement of US Groups in the TDAQ}
The US has played a strong role in the development and operation of
the ATLAS TDAQ system. US groups are making major contributions to
TDAQ effort in the following areas:
\begin{itemize}
  \item Calorimeter and Muon sub-system electronics that provide data
    to the L1 trigger system;
  \item L1 Calorimeter Trigger upgrades in \PhZ\ and \PhI ;
  \item the L2 Fast Tracker (FTK) trigger;
  \item the Region-of-Interest Builder;
  \item development of general-purpose, ATCA-based RODs;
  \item HLT and DAQ core software, particularly in the areas of
    dataflow and parallelization.
\end{itemize}

\subsection{Phase-II Trigger Architecture}
\label{sect:tdaq:arch}

\subsubsection{Rate Estimates for the Trigger at \PhII\ Luminosities}
The main issues in continuing to operate the \PhI\ trigger
system at luminosities up to \PhIILumiMax\ (a factor of 2--3 higher than
those foreseen in \PhI ) come from the L1 system. The \PhI\ L1 trigger
is constrained to operate with a maximum latency of 2.5 \us\ and a
maximum accept rate of 100 kHz due to detector readout
capability. This limits possibilities for coping with higher rates
expected in \PhII .

Estimates of the rates of \PhI\ L1 leptons triggers have been made
based on simulations of the functionality
of the \PhI\ electromagnetic calorimeter Feature Extractors and 
extrapolations of the performance of the muon trigger using existing
data, including the expected performance of the new forward muon
chambers, the New Small Wheels.
These estimates indicate that a single electron/photon
trigger with an \ET\ threshold of 25 GeV including hadronic isolation
requirements would produce a L1 rate of 125 kHz. To achieve a
manageable rate of 20 kHz would require raising this threshold to 40
GeV. For single muon triggers with a \pT\ of 20 GeV rates above 40 kHz
are predicted, depending upon assumptions made about background
conditions. In both of these cases, ATLAS physics goals would be
severely compromised by L1 trigger restrictions. 
Estimates of expected \PhI L1 trigger rates for a variety of different
triggers at \PhIILumiMax\ are given in
Table~\ref{table:tdaq:rates}~\cite{Ph2-LoI}.

\begin{table}[htbp]
\begin{center}
\caption{Expected trigger rates at \PhIILumiMax\
  for the L1 \PhI\ and the baseline split
  L0/L1 \PhII trigger systems. 
  The EM triggers all assume the hadronic energy
  veto (VH) is used. 
  For the photon and di-photon triggers it is
  assumed that the full granularity in the L1 calorimeter trigger
  will bring an additional factor 3 in background rejection power. The
  $\tau\tau$ trigger rate assumes a factor 2 reduction in the tau fake
  rate from the eFeX. The {\it exclusive} rates for $e\tau$ and
  $\mu\tau$ are not included as these will depend strongly on the
  exact trigger menu and trigger thresholds used.
  The rates for the JET and MET triggers are
  estimates based on an extrapolation of the current
  fraction of the trigger budget used for these triggers.  
\label{table:tdaq:rates}}
\begin{tabular}{llcc}
\hline
  {\bf Object(s)} & {\bf Trigger} & \multicolumn{2}{c}{\bf Estimated Rate}\\
  & &  \PhI L1 / \PhII  L0 & \PhII L1\\ 
\hline        
  $e$             &  {\tt EM20}      &       200\,kHz &       40\,kHz\\
  $\gamma$        &  {\tt EM40}      &        20\,kHz &       10\,kHz\\
  $\mu$           &  {\tt MU20}      &     $>$40\,kHz &       10\,kHz\\
  $\tau$          &  {\tt TAU50}     &        50\,kHz &        20\,kHz\\
\hline
  $ee$            &  {\tt 2EM10}     &        40\,kHz &      $<$1\,kHz\\
  $\gamma\gamma$  &  {\tt 2EM10}     &       as above &   $\sim$5\,kHz\\
  $e\mu$          &  {\tt EM10\_MU6} &        30\,kHz &      $<$1\,kHz\\
  $\mu\mu$        &   {\tt 2MU10}    &         4\,kHz &      $<$1\,kHz\\
  $\tau\tau$      &   {\tt 2TAU15I}  &        40\,kHz &         2\,kHz\\
  Other           &  {\tt JET + MET} & $\sim$100\,kHz & $\sim$100\,kHz\\
\hline
  Total           &                  & $\sim$500\,kHz & $\sim$200\,kHz\\
\hline
\end{tabular}
\end{center}
\end{table}

\subsubsection{Proposed Phase-II Architecture}
The architecture of the \PhII\ trigger system is driven by the ATLAS
physics goals described in Ref.~\cite{SnowmassPhysWP} and by constraints
imposed by the detector sub-systems. As mentioned above, one of the
most stringent physics requirements is the need to maintain efficient
L1 single-lepton triggers with thresholds in the 20 GeV range. On the
detector side, planned \PhII\ upgrades relax the L1 rate and latency
limitations from their \PhI\ values of 100 kHz and 2.5 \us . 
However, the inaccessibility of approximately 30\% of the electronics
of the MDT system's Barrel Inner (BI) layer results in this system
being the new bottleneck. The existing MDT electronics can operate
with a latency of up to 20 \us\ at a maximum L1 accept rate of
$\sim$200 kHz. 

As can be seen from Table~\ref{table:tdaq:rates}, the total rate for
L1 triggers that allow ATLAS to achieve its \PhII\ physics goals using
the \PhI\ hardware exceeds the 200 kHz maximum by a factor of 2.5. 
The strategy proposed by ATLAS to deal with this involves creating a
two-stage hardware trigger. Level-0 (L0), which will use
the \PhI\ L1 trigger hardware, accepts events at a rate of 500 kHz
with a latency of 6 \us . Following an L0 accept a new L1 system,
using additional information and more sophisticated algorithms, will
reduce the accept rate to the 200 kHz target with an added latency of
14 \us . Thus, the total latency from the L0/L1 system matches the
20 \us\ constraint from the MDT readout electronics.
Four new features at L1, made possible by changes to the ATLAS
detector in \PhII , are foreseen to make possible this extra rate
reduction.
\begin{enumerate}
  \item {\bf Tracking Information:} 
    Association of tracks found using data from the silicon inner
    tracker (ITK) to calorimeter
    objects and to muons will provide a substantial reduction in the
    electron, muon, and tau trigger rates.
  \item {\bf Full Granularity Calorimeter Information:} 
    All data from both the LAr and TileCal calorimeters will be
    available to the L1Calo trigger providing extra background
    rejection in (di)photon and (di)tau triggers.
  \item {\bf MDT Information:}
    Muon track momentum reconstruction, using the precise Monitored
    Drift Tubes (MDT), will be possible at L1. This is particularly
    effective at rejecting background for relatively low momentum
    muons. 
  \item {\bf Topology:}
    The \PhI\ L1Topo trigger will be enhanced to include new
    information from L1Track, L1Calo, and L1Muon.
\end{enumerate}

\begin{figure}[htbp]
\begin{center}
  \includegraphics[width=0.75\textwidth]{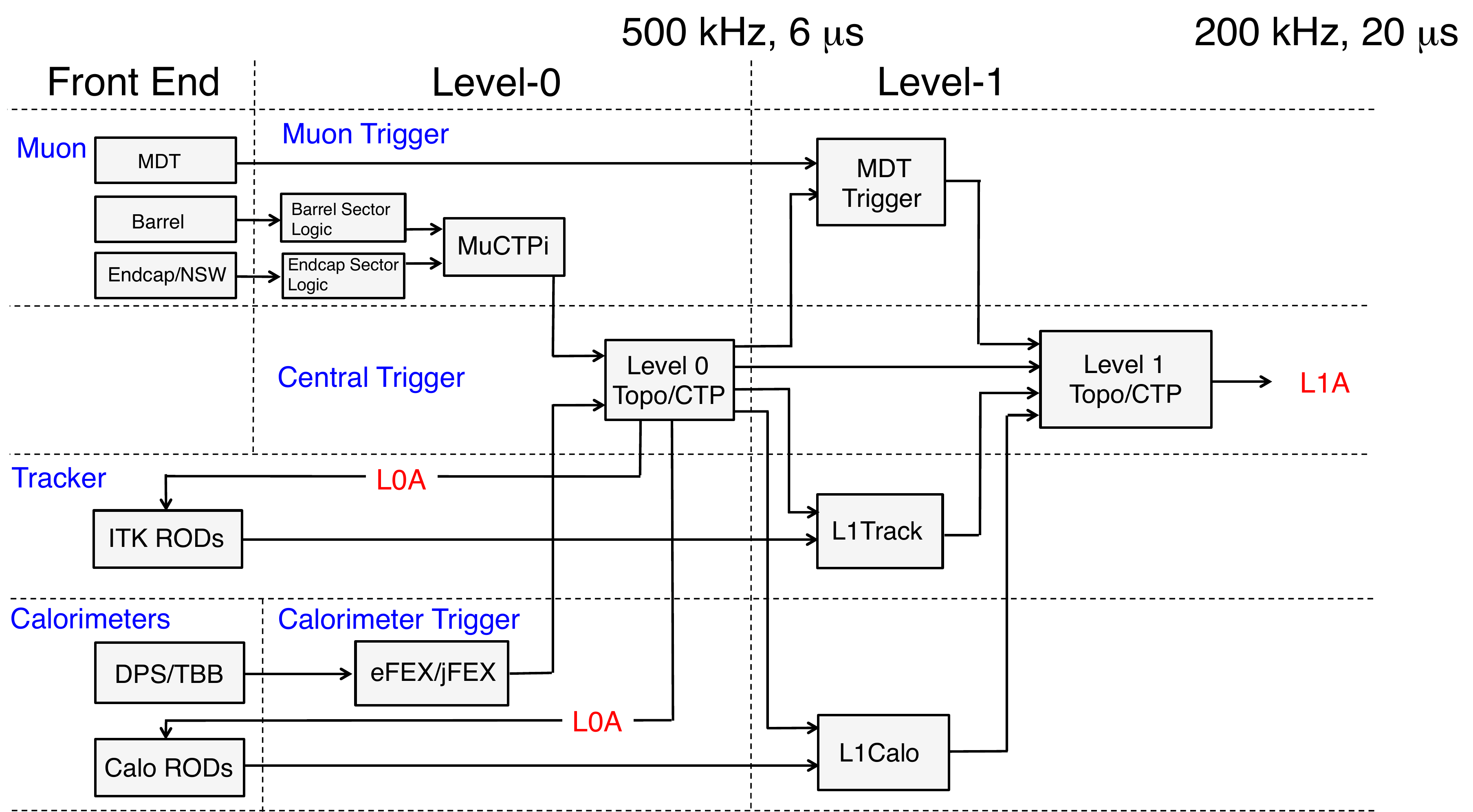}
  \caption{
    Functional diagram of the L0/L1 trigger system
    in \PhII .
  }
  \label{fig:tdaq:PhII}
\end{center}
\end{figure}

The proposed \PhII\ L0/L1 trigger architecture is shown in
Figure~\ref{fig:tdaq:PhII}. Its effect on estimated trigger rates is
summarized in Table~\ref{table:tdaq:rates}. More information on the
individual elements of the system is given in the following and
in~\cite{Ph2-LoI}. 

\subsection{Calorimeter Trigger}
\label{sect:tdaq:calo}
As described in Section~\ref{sect:calo}, the entire calorimeter
front-end and back-end electronics systems will be replaced in \PhII ,
allowing digitization of all channels on every bunch crossing and
transmission of this data off-detector.
The low-level calorimeter trigger in \PhII\ 
will take advantage of this new, higher-granularity data
and will be separated into two pieces. 
Level-0 (L0Calo) will approximate the functionality of
the \PhI\ L1Calo but will produce accepts at a much higher rate.
Level-1 (L1Calo) will use the full-granularity calorimeter
information and more sophisticated algorithms to process events at the
average L0 accept rate of 500 kHz, up to a possible peak rate of 20
MHz.

The \PhII\ L0Calo system will use electron and jet feature extractors
(FEXs) as in the \PhI\ L1Calo. However, L0Calo inputs will differ from
those in \PhI\ due to higher precision data being available at the
trigger level in the new calorimeter readout scheme. 
As in \PhI , HCAL data will be
divided into regions of size \AxB{0.1}{0.1} in \AxB{$\eta$}{$\phi$}, while
four layers of LAr ECAL data (pre-sampler and three ECAL sampling
layers) will be delivered in the ``1-4-4-1'' arrangement. In this
arrangement, the first two ECAL sampling layers have a four times
finer segmentation in $\eta$ (\AxB{0.025}{0.1} in \AxB{$\eta$}{$\phi$})
than the pre-sampler and third ECAL sampling layers. All of this data
is sent to the L0Calo on 4064, 10 Gb/s optical links, where it will be
processed in the FEXs using firmware modified from the \PhI\ versions,
but running similar algorithms to \PhI\ L1Calo.

The \PhII\ L1Calo system will have access to full granularity
calorimeter data in Regions of Interest (ROIs) defined by L0 and will
therefore be able to deliver improved measurements of the energies and
positions of trigger objects. For example, the second ECAL sampling
layer has a full granularity segmentation of \AxB{0.025}{0.025}
in \AxB{$\eta$}{$\phi$} (compared to \AxB{0.025}{0.1} at L0Calo),
while the very finely grained first ECAL sampling layer
(\AxB{0.003125}{0.1} in \AxB{$\eta$}{$\phi$}) could be used to suppress EM
trigger backgrounds due to $\pi^0 \to \gamma \gamma$ decays.

A common area of \RD\ in both of these systems centers on the large
data volumes (2 Tbps in L0Calo and 200 Tbps in L1Calo) sent to and
among system elements. 

\subsection{Muon Trigger}
\label{sect:tdaq:muon}
Upgrades to the \PhI\ L1Muon system required to retain \pT\ thresholds
in the 20 GeV range for the split L0/L1 trigger scheme in \PhII\ fall
into three main categories.
First, most muon readout electronics in the barrel and endcaps will
need to be replaced to deal with the new trigger architecture.
Second, tracking performance of the Resistive Plate Chambers (RPCs) in
the barrel can be improved by using time-over-threshold information.
Finally, precise hit information from the Monitored Drift Tubes (MDTs)
will be added to the L0 or L1 trigger logic to improve muon trigger
momentum resolution, aiding in the rejection of low-momentum
backgrounds.

For the MDTs, bunch-crossing identification for hits in individual
tubes will be transmitted to the L0 or L1 trigger systems using high
speed optical links. This information will allow \pT\ reconstruction
at L0 or L1 with a quality similar to that of the current L2 muons,
resulting in a reduction of the L1Muon rate by a factor of
approximately three over much of the detector.

The L1Muon \pT\ resolution can also be sharpened by using the charge
distribution in clusters of RPC $\eta$ strips. 
This charge distribution can be accessed because the signal duration
in the current RPC front-end electronics is correlated to the input
charge. Measurements with a TDC could thus deliver the charge
distribution in adjacent strips, from which a centroid could be
formed.
\RD\ is currently under way to evaluate the potential of this idea
in detail.

\subsection{Level-1 Track Trigger}
\label{sect:tdaq:l1track}
\begin{figure}[htbp]
\begin{center}
  \subfigure[]{
    \label{fig:tdaq:L1TrMu}
    \includegraphics[width=0.45\textwidth]{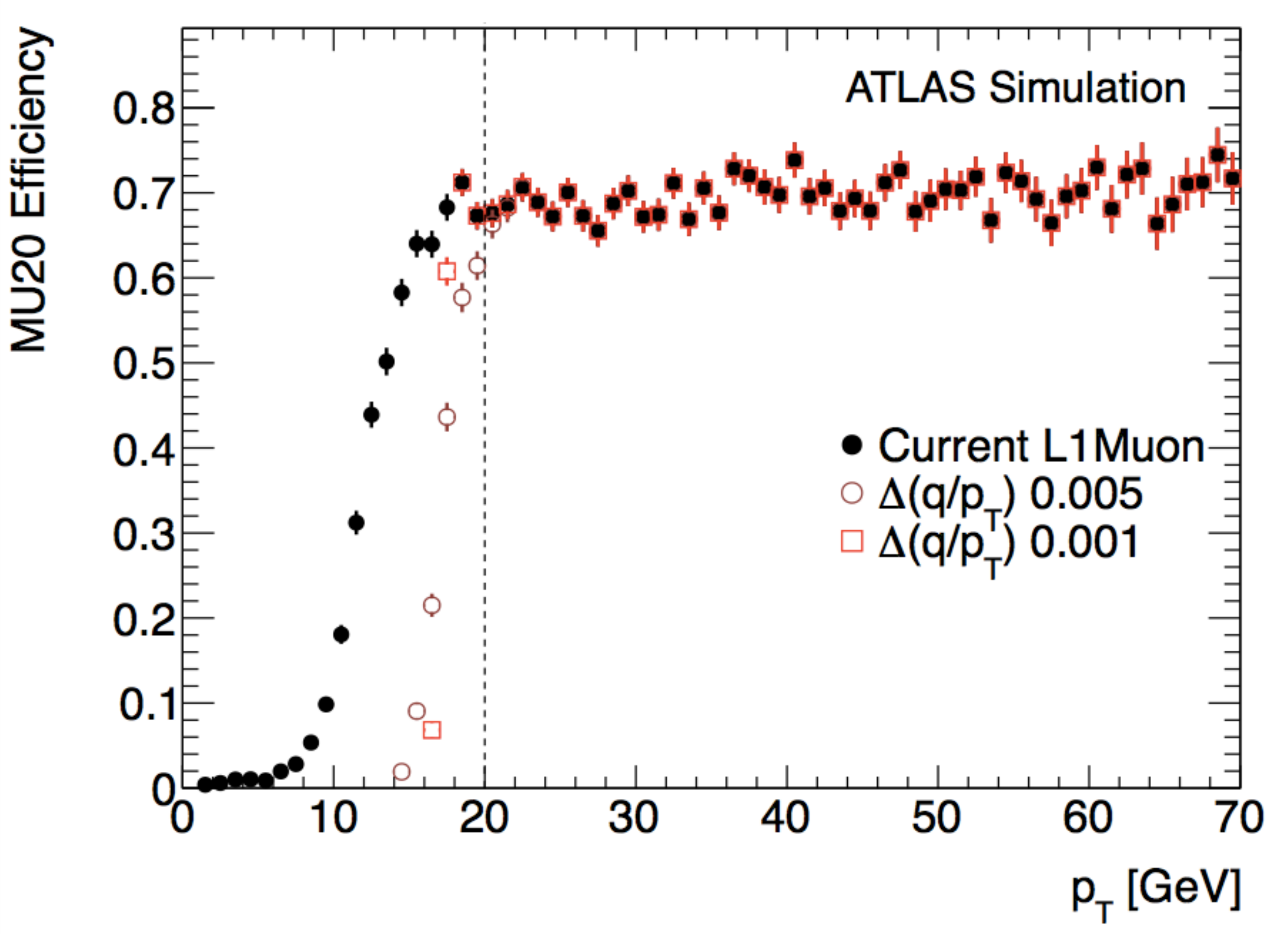}}
  \subfigure[]{
    \label{fig:tdaq:L1TrE}
    \includegraphics[width=0.45\textwidth]{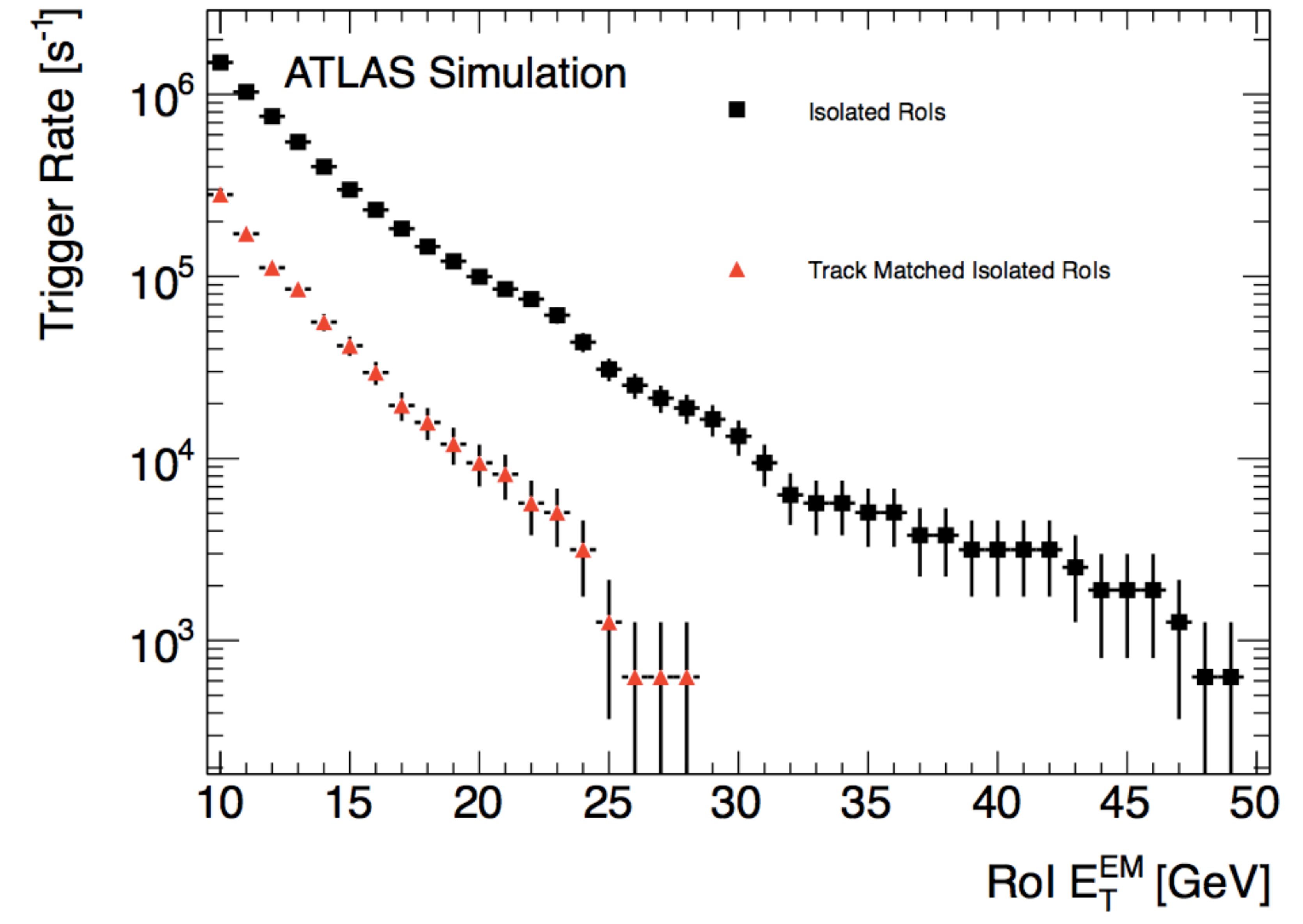}}
  \caption{
    \subref{fig:tdaq:L1TrMu} Muon trigger efficiencies with a 20 GeV
    threshold as a function of true muon \pT\ after matching to a true
    muon, assuming a track trigger with different \pT\ resolutions.
    \subref{fig:tdaq:L1TrE} Trigger rate vs L1 EM cluster \ET\
    threshold for simulated minimum bias events with \aveMu\ = 70.
  }
  \label{fig:tdaq:L1Track}
\end{center}
\end{figure}

Despite improvements to L1Calo and L1Muon in \PhII ,
these systems alone are unlikely to entirely meet the physics goals of
ATLAS. For this reason a crucial element of the ATLAS \PhII\ upgrade
program is the addition of a new trigger system at L1 using
information from the Inner Detector. Charged particle tracks
reconstructed by this L1Track trigger would be combined with EM
objects from L1Calo and muon candidates from L1Muon to yield
significant rate reductions. Current estimates, using simulation
studies and extrapolations of current L1 trigger rates from data,
indicate that reductions in the rate of more than a factor of five
compared to L0 for
20 GeV single muon triggers and 18 GeV isolated single electron
triggers using L1Track information are possible. 
Examples of these studies are give in Figure~\ref{fig:tdaq:L1Track}. 

Two possible L1Track architectures are currently under study. The
baseline architecture, {\it RoI-Driven L1Track}, would provide all
tracks in areas near L0 EM and Muon RoIs, 
while an alternate, {\it Self-Seeded L1Track}, would reconstruct all
high momentum (\pT\ $>$ 10 GeV) tracks independent of L0 RoIs.

In the RoI-Driven approach, Inner Detector data, buffered in
on-detector electronics, is read out through associated Readout
Drivers (RODs) only for those regions near a L0
EM or Muon RoI. This regional readout feature is already included in
the design of \PhII\ front-end readout ASICs for the strip tracker
upgrade. The main challenge will be to perform the data readout and
tracking within the \PhII\ L1 latency budget of 20 \us . Preliminary
studies indicate that this will be feasible.

The alternate, Self-Seeded, approach to L1Track requires a huge
reduction in inner detector data volume.
This could be accomplished by a combination of
the use of fewer tracking layers in the trigger,
and the early rejection of low \pT\ tracks using cluster sizes and the
inclination angle between the hits in stacked double strip layers.
This approach requires major changes to the layout of the inner
detector layers over what is currently assumed (see
Section~\ref{sect:track}).

In both approaches L1Track pattern recognition might be done using
associative memory (AM) technology similar to that being developed for
the \PhI\ ATLAS L2 Fast Tracker (FTK)~\cite{Ph1-LoI}. \RD\ is ongoing
to exploit novel ASIC design technologies, such as 3-D, to meet the
experiment's goals here.

\subsection{Central Trigger System}
\label{sect:tdaq:ctp}
Changes to the Central Trigger (CT) for \PhII\ include development of
separate L0 and L1 CTs, with topological capabilities included in
each; and an update of the trigger, timing, and control (TTC)
system. As can be seen from Figure~\ref{fig:tdaq:PhII}, the L0CT and
L1CT are structurally rather similar. The intent is to use the same
technologies for these systems to ease design and maintenance
requirements. For the upgraded TTC, the required topology matches that
of a Passive Optical Network (PON). Studies are under way to assess
the feasibility of using commercially available
components for the final system.

\subsection{High Level Trigger}
\label{sect:tdaq:hlt}
In order to remain within limitations of 5--10 kHz on the total data
recording rate, the \PhII\ High Level Trigger (HLT) will need to
provide a rejection factor of 20--40 beyond the L1 accept rate of 200
kHz. To achieve this goal, the HLT will employ offline-type selections
and will rely increasingly on multi-object signatures. Advances in
computer hardware performance over the next 10 years will help here,
but upgrades to the HLT farm as well as improvements to selection
software will also be required. One particularly promising area of
development in HLT software involves the increased use of many-core
architectures (e.g.~GPUs) and parallelization of code. This will
require significant changes to both the HLT framework and to the
algorithms themselves. Another challenge will be to maintain
commonality with offline software, a feature that substantially eases
the burden of code development and maintenance.

\subsection{Data Acquisition}
\label{sect:tdaq:daq}
\begin{figure}[htbp]
\begin{center}
  \includegraphics[width=0.75\textwidth]{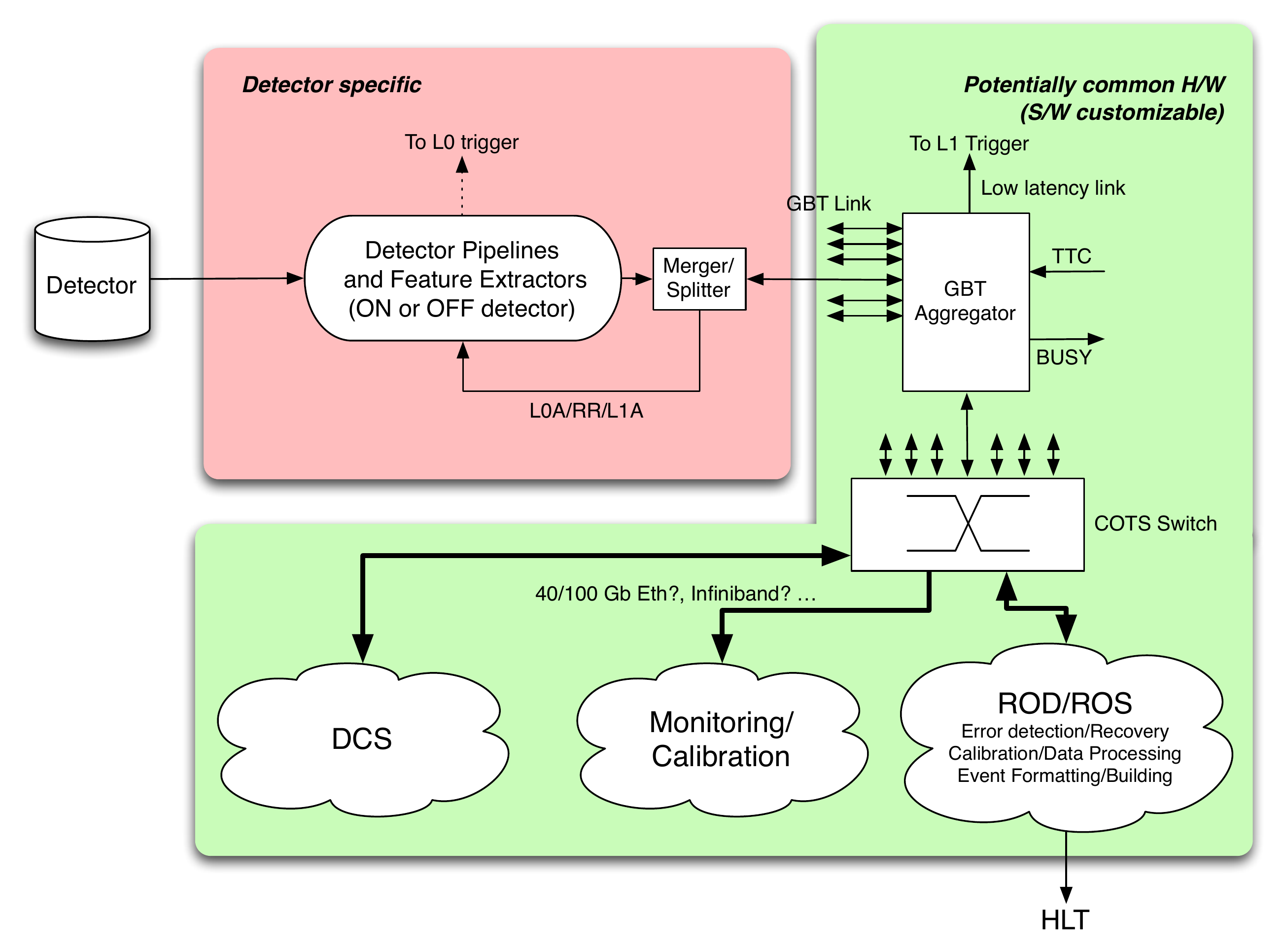}
  \caption{
    Readout architecture overview for the \PhII\ upgrade.
  }
  \label{fig:tdaq:DaqPhII}
\end{center}
\end{figure}

Increased L1 accept rates (200 kHz) and larger event sizes ($>$4 MB)
due to high levels of pileup at \PhII\ luminosities indicate that at
least a factor of four increase in bandwidth will be required of the
ATLAS data acquisition system in \PhII\ beyond that needed in \PhI\
running. Additionally, higher values of pileup (\aveMu\ approaching
200 at luminosities of \PhIILumiMax ) imply a significant increase in
event processing time. Meeting these challenges will require changes
in both hardware and software. Principal areas under study are:
network technologies,
online databases,
information sharing mechanisms,
and expert systems.
Although it is premature to specify details of an upgraded data
acquisition system, given the speed at which the relevant technologies
are developing, a possible architecture for \PhII\ readout is given in
Figure~\ref{fig:tdaq:DaqPhII}. 
Figure~\ref{fig:tdaq:PhI}~\subref{fig:tdaq:DaqPhI}
shows the current readout architecture for comparison.
The \PhII\ system aims for increased levels of commonality across
detector sub-systems, taking advantage of emerging technologies and
commercially available components. For example, aggregators
feeding commercial, high-speed network switches allow data from
different sub-systems to be read out into a common Readout Driver
(ROD). Aside from simplifying issues of production and maintenance,
this scheme could allow the possibility of re-organizing readout
connectivity without physical re-cabling as well as leading to
increased flexibility in terms of scalability and staging.

\subsection{Summary of Main TDAQ R\&D Areas}
\label{sect:tdaq:rd}
Aside from system-specific studies, the main areas of
TDAQ-related \RD\ that are of general interest are summarized below.
Work in most of these areas is just starting. 
However, most of the approximately 15 US institutes currently
participating in TDAQ activities
are playing strong roles in upgrade \RD\ planning.

\begin{enumerate}
  \item {\bf High-speed optical links:} 
    Particular areas of concentration here are in the Calorimeter
    trigger system where
    approximately 200 Tbits of data must be
    transmitted per second in L1Calo;
    and in detector front-end to back-end links, which must operate at
    speeds of up to 10 Gbps in high radiation environments.
  \item {\bf Novel ASIC technologies:} 
    For example, 3D technologies for use in Track Trigger Associative
    Memory chips.
  \item {\bf ATCA applications:} 
    Standardized RODs implemented as ATCA cards could save costs by
    providing high performance, flexible platforms for a wide variety
    of readout applications.
  \item {\bf Parallelism and multi-core processing applications:}
    \RD\ is ongoing in using multi-threaded code and GPU architectures
    to improve performance and open new possibilities in the HLT and
    DAQ.
  \item {\bf High-speed switching networks:}
    Clever use of commercial high-speed network switches within a
    common readout framework could provide a flexible, cost-effective
    solution to sub-system readout needs.
\end{enumerate}

\section{Conclusion}
\label{sect:concl}
The ATLAS \PhII\ upgrade will allow exploration of the Energy
Frontier to the highest masses as well as precision studies of the
Higgs boson and any other new phenomena, fulfilling the promise of the
LHC.  To achieve this requires significant upgrades to the detector
focused primarily on a new tracking detector, new calorimeter
electronics, and a new TDAQ system.
All of these upgrades to ATLAS require
near-term \RD\ on sensors, electronics, and high bandwidth
data-transmission components, areas in which US groups are already playing
leading roles. 
Upgrades are also planned to the ATLAS muon system, particularly
to the readout electronics and trigger systems.
US groups have broad expertise in these areas and could contribute
effectively to the \PhII\ muon system upgrade effort were more
\RD\ money to become available.
Although the focus of this whitepaper has been on
ATLAS and the LHC, many of the advances planned will be of general
value to the community as they require mastering the latest
technologies and exploring their limits.

\clearpage
\bibliographystyle{unsrt}
\bibliography{atlas-instrum}
\end{document}